\documentclass[%
 reprint,
 amsmath,amssymb,
 aps,
]{revtex4-2}

\usepackage{dcolumn}
\usepackage{bm}
\usepackage{amsfonts}
\usepackage{natbib}
\usepackage{bibentry}
\usepackage[table]{xcolor}
\usepackage{graphicx}
\usepackage{epstopdf}
\usepackage{amssymb}
\usepackage{amsmath}
\usepackage{dcolumn}
\usepackage{subfigure}
\usepackage{siunitx}
\usepackage{mathptmx}

\DeclareGraphicsExtensions{.eps,.png,.pdf}

\usepackage{hyperref}
\hypersetup{
colorlinks   = true, 
urlcolor     = blue, 
linkcolor    = blue, 
citecolor   = blue 
}
\usepackage{setspace}
\usepackage{enumitem}
\usepackage{fancybox}
\usepackage{comment}
\usepackage{url}
\usepackage{here}
\setlist{nolistsep}
\usepackage{multimedia}
\usepackage{soul}
\setstcolor{red}
\DeclareMathAlphabet      {\mathbf}{OT1}{cmr}{bx}{n}

\usepackage{siunitx}
\begin{document}

\preprint{APS/123-QED}

\title{Acousto-microfluidic Control of Liquid Crystals}

\author{Gustavo A. V\'asquez-Montoya$^{1}$}
\thanks{These authors contributed equally to this work.}
\author{Tadej Emer\v{s}i\v{c}$^{1}$}
\thanks{These authors contributed equally to this work.}
\author{Noe Atzin$^1$}
\author{Antonio Tavera-V\'azquez$^1$}
\author{Ali Mozaffari$^{1,2}$}
\author{Rui Zhang$^3$}
\author{Orlando Guzm\'an$^4$}
\author{Alexey Snezhko$^5$}
\author{Paul F. Nealey$^{1,5}$}
\author{Juan J. de Pablo$^{1,5}$}
\altaffiliation{Corresponding author. E-mail: depablo@uchicago.edu (J.J.d.P.)}

\affiliation{\vspace{4pt}$^1$Pritzker School of Molecular Engineering, The University of Chicago, Chicago, IL 60637, USA\\$^2$OpenEye Scientific, Cadence Molecular Sciences, Boston, Massachusetts 02114, USA\\$^3$Department of Physics, Hong Kong University of Science and Technology, Clear Water Bay, Kowloon, Hong Kong\\$^4$Departamento de F\'isica, Universidad Aut\'onoma Metropolitana Iztapalapa, Av. San Rafael Atlixco 186, Ciudad de M\'exico 09340, Mexico\\$^5$Materials Science Division, Argonne National Laboratory, Lemont, IL 60439, USA} 

\begin{abstract}
The optical properties of liquid crystals serve as the basis for display, diagnostic, and sensing technologies. Such properties are generally controlled by relying on electric fields. In this work, we investigate the effects of microfluidic flows and acoustic fields on the molecular orientation and the corresponding optical response of nematic liquid crystals. Several previously unknown structures are identified, which are rationalized in terms of a state diagram as a function of the strengths of the flow and the acoustic field. The new structures are interpreted by relying on calculations with a free energy functional expressed in terms of the tensorial order parameter, using continuum theory simulations in the Landau-de Gennes framework. Taken together, the findings presented here offer promise for the development of new systems based on combinations of sound, flow, and confinement.
\end{abstract}


\maketitle

\section{\label{sec:level1}INTRODUCTION}
Optofluidic devices, which combine the transport features of liquids and the remote addressability offered by optical properties, offer considerable potential applications in sensing and display technologies \cite{Psaltis06,Monat,Yang09,Psaltis11}. In particular, liquid crystals (LCs), which exhibit versatile and anisotropic optical properties associated with their alignment, provide attractive platforms for optofluidic applications \cite{Psaltis1111,Zhao19,Gibbons91,Schadt96}. LCs are highly sensitive to external stimuli, including electric and optical fields, and have been widely used in displays and photonic systems \cite{Reinitzer88,Khoo,Schiekel71}. Recently, advances in nanofabrication techniques have created new opportunities for the combined application of multiple fields, such as ultrasound and flow.

Acoustic fields are typically used in miniaturized devices in the form of bulk acoustic waves or surface acoustic waves (SAWs). Here we focus on SAWs, which exhibit enhanced sensitivity compared to bulk wave devices \cite{ding13}. The frequency of SAWs ranges from several hundred MHz to a few GHz; they have been considered in applications that include radio-electronic components and sensors \cite{Hashimoto00,Liu16}. Recently, techniques that rely on standing SAWs (SSAWs) formed by two opposite and coaxial waves have also been used for manipulation of biological cells and microparticles, serving to highlight their potential for manipulation of soft materials \cite{Guo16,Ding12,Rapp13,Zhang14,Chen14,Ding14}. Past studies of the effects of SAWs on nematic LCs (NLCs) have considered periodic pressure fields, acoustic streaming flows, and other applied fields \cite{Sliw83,Fang77,Sato81,Toda02,Lee01,Song18}. It has been shown that, for perpendicular (homeotropically) anchored nematics, SAWs lead to the formation of stripe patterns that gradually transition into a dynamic scattering regime characterized by a turbulent-like flow behavior, where the nematic director orientation is randomized. While a variety of demonstrations have focused on tuning transparency and light scattering in cholesteric LCs, polymer-dispersed LC screens, acoustic images visualized on LCs, medical imaging, and LC tunable lenses \cite{Toda04,Ozaki08,Liu11,Gerdt99,Sandhu01,Sandhu09,Shimizu18}, a fundamental understanding of the interaction between non-equilibrium structures and SAWs in LCs is still missing. In this work, we present a systematic theoretical and experimental study of the structure of NLCs under the influence of combined acoustic and flow fields.

Applied flows are known to have a strong effect on the orientation and order of NLCs \cite{sengupta13}. Recent studies have revealed the existence of different topological states in channel-confined nematic flows — referred to as bowser, chiral, and dowser states \cite{copar20}. Experiments with laser tweezers have shown that these topological states can also be locally created and controlled \cite{emersic19}. Nevertheless, the effect of an acoustic field on topological states in channel-confined nematic flows remained unexplored.

The optofluidic system considered in this work consists of a NLC confined by a PDMS microfluidic channel that is coupled to a SSAW generator. We show that different structures arise in the LC depending on the acoustic and flow field intensities. First, we characterize optical patterns driven only by the SSAWs. We rely on polarized optical microscopy (POM) and fluorescence confocal polarizing microscopy (FCPM) to visualize the reorientation of the LC mesogens. We also characterize the temperature changes in the system in terms of acoustic strength and determine the level of influence on the observed optical transitions. After identifying the characteristics of the LC under acoustic fields, we add microfluidic flows and examine the response of the LC, thereby producing a state diagram of optical texture in terms of relevant non-dimensional parameters such as the streaming Reynolds number and the Ericksen number. The experimental results are interpreted using continuum simulations with a Landau-de Gennes (LdG) free energy functional for the tensor order parameter. By combining simulations and experiments, this work provides a detailed picture of the transition between different structures and helps provide a foundation for design of LC-based optofluidic devices controlled by acoustic waves.

\section{IDENTIFICATION OF ACOUSTICALLY INDUCED STRUCTURES OF CONFINED NLC}
\begin{figure}[h]
\includegraphics{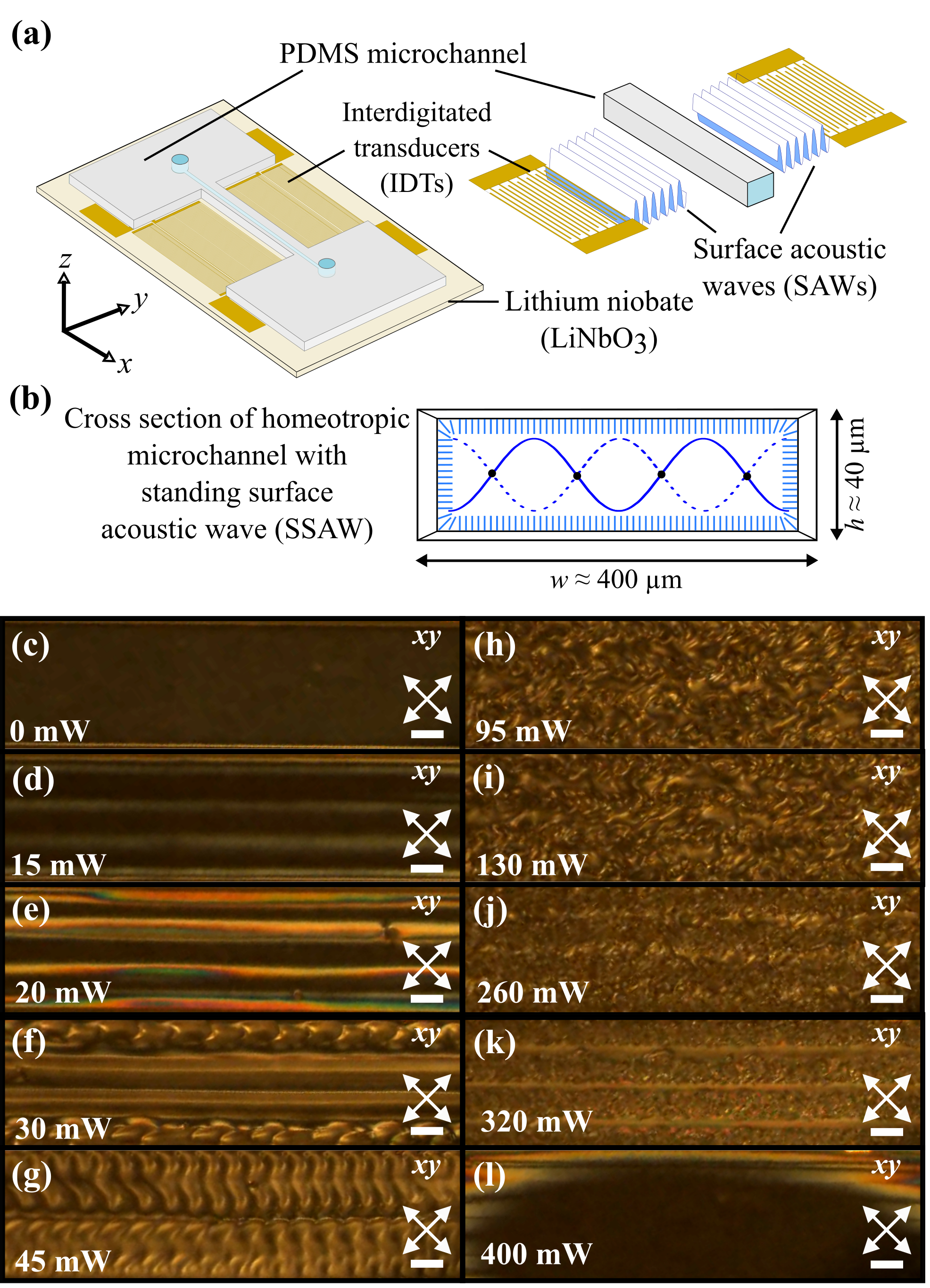}
\caption{\label{Fig1} NLC in homeotropic microfluidic channel under SSAWs. (a) Schematic representation of PDMS channel on a lithium niobate substrate (LiNbO$_3$) with two pairs of parallel IDTs that generate SSAWs. (b) Cross section of the channel of size $400 \times 40$ \SI{}{\micro\metre} with the nodes and anti-nodes of the standing pressure wave with wavelength $200$ \SI{}{\micro\metre}. (c)-(l) Experimental POM images show a top view of the NLC in a channel under SSAWs. Different values of the input power applied to the IDTs lead to different structures: (c) no patterns, (d) white stripe patterns, (e) color stripe patterns, (f) brown stripe patterns, (g) dynamical behavior of spatially periodic patterns, (h)-(i) dynamic scattering characterized by a turbulent-like flow behavior, (j)-(k) dynamic scattering with stripes  and (l) isotropic phase transition. White crossed double arrows show the orientation of the polarizers. Scale bars are $100$ \SI{}{\micro\metre}.  }
\end{figure}

All experiments are performed on 5CB confined in a linear microfluidic channel. The channel has a rectangular cross section, with height $h=$ \SI{40}{\micro\metre} and width $w=$ \SI{400}{\micro\metre}, and is fabricated out of a PDMS (polydimethylsiloxane) relief bounded to a piezoelectric lithium niobate (LiNbO$_3$) substrate [Fig.~\ref{Fig1}(a)]. The NLC is loaded into the channel after treating it with DMOAP to achieve homeotropic surface alignment [Fig.~\ref{Fig1}(b)]. As shown in Fig.~\ref{Fig1}(a), the channel is centrally positioned between two parallel interdigitated transducers (IDTs) patterned on the piezoelectric substrate. A radio frequency (RF) signal from a signal generator is applied to the IDTs to convert the electric signal into SAWs that propagate on the substrate surface in the $y$-direction. The superposition of two counter-propagating SAWs results in a SSAW with wavelength around \SI{200}{\micro\metre} [Fig. \ref{Fig1}(b)], determined by the IDT pitch. The SSAW transmitted through the channel filled with the NLC is a standing pressure wave with 4 nodes, as illustrated in Fig.~\ref{Fig1}(b).

\begin{figure*}[ht]
\includegraphics{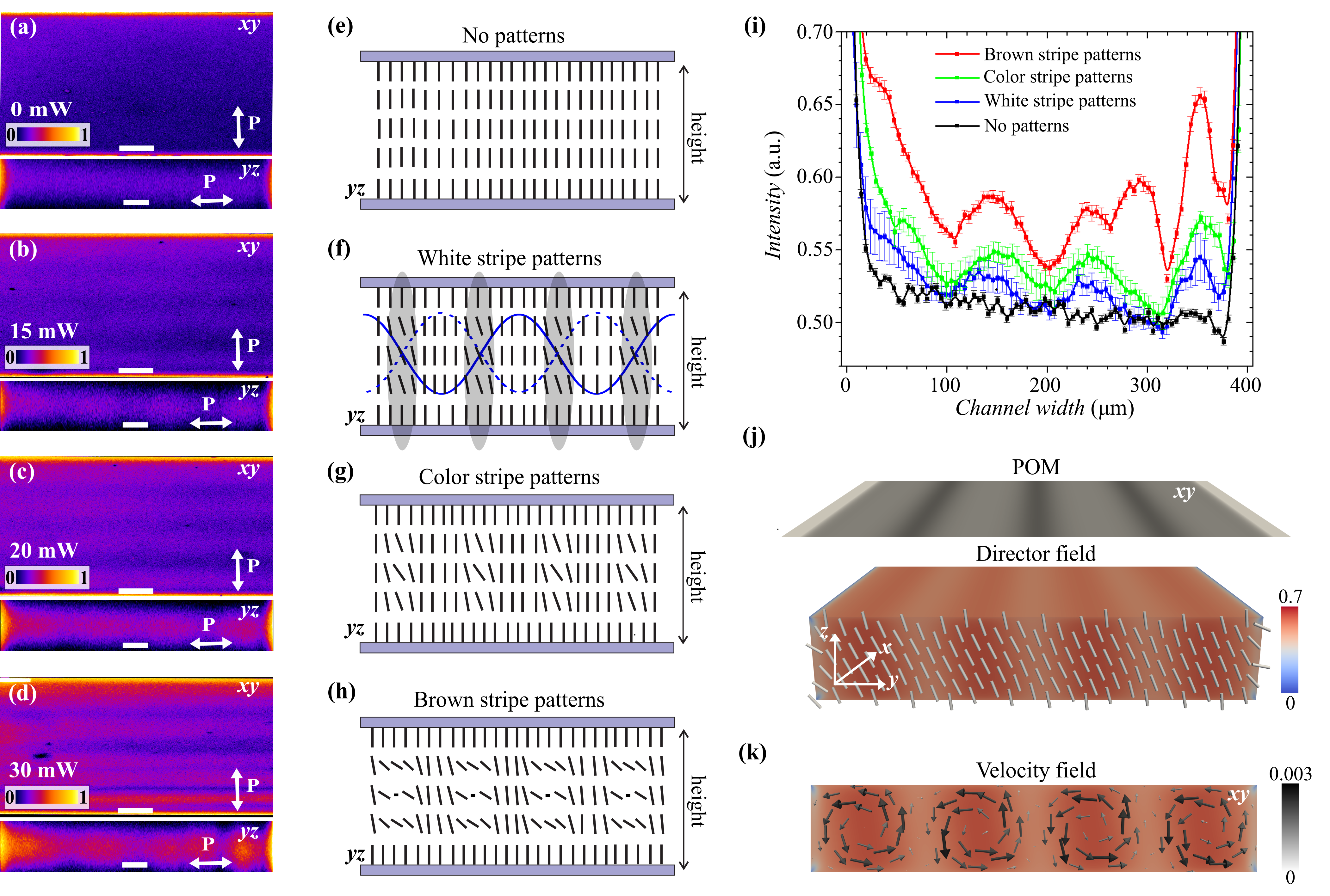}
\caption{\label{Fig2} Formation of stripe patterns. (a)-(d) FCPM top view (\textit{xy} plane) and cross section (\textit{yz} plane) of a microfluidic channel, corresponding to NLC in (a) the absence of patterns, and in (b) the presence of white, (c) color, and (d) brown stripe patterns, as described in Fig.~\ref{Fig1}. Scale bars are \SI{100}{\micro\metre} (\textit{xy} plane) and \SI{20}{\micro\metre} (\textit{yz} plane). (e)-(h) Corresponding schematic representation of the director orientation across the channel cross section for stripe patterns in (e) the absence of patterns, and in (f) the presence of white, (g) color, and (h) brown stripe patterns. The grey area in (f) indicates the pressure node regions tilting the NLC molecules. Side walls are not included. (i) Fluorescent signal intensity as a function of channel width for all stripe patterns. (j) Cross polarized image of stripe patterns with the corresponding director field orientation [see also Fig.~\ref{FigS3}] and scalar order parameter (see color bar) predicted by numerical simulations. Tilted molecules at the pressure nodes result in a periodic distribution of the order parameter across the channel. (k) Numerically predicted acoustic streaming flow (represented with black/white arrows, see color bar) around the pressure nodes. }
\end{figure*}

The acoustically induced structures of NLC in the microfluidic channel are observed under POM [Figs.~\ref{Fig1}(c)-\ref{Fig1}(l)]. Without an acoustic field, the channel appears dark, corresponding to a uniformly aligned director field along the $z$-axis [Fig.~\ref{Fig1}(c)]. Application of the acoustic field leads to structure formation in the NLC. With an applied input power of the RF signal of \SI{15}{\milli\watt}, the SSAW induces the formation of stripe patterns perpendicular to the sound propagation direction located at the SSAW pressure nodes [white stripes, Fig.~\ref{Fig1}(d)]. The pattern undergoes a transition when the input power increases to \SI{20}{\milli\watt} characterized by the appearance of colors in the stripes due to birefringence [color stripes, Fig.~\ref{Fig1}(e)]. A further increase of the power induces a discontinuous transition of the patterns where the birefringent colors are replaced in favor of broader stripes with lower intensity of transmitted light [brown stripes, Fig.~\ref{Fig1}(f)]. This transition is similar to the homeotropic-dowser transition typically observed in homeotropic nematic samples \cite{pieranski16}. In addition, a disruption of the texture is observed near the channel walls, similar to streaming-induced rolls. Around an input power of \SI{45}{\milli\watt}, the stripes transition into a dynamical behavior consisting of spatially periodic patterns, as seen in Fig.~\ref{Fig1}(g). Higher acoustic intensity leads to a disorganized turbulent-like flow behavior assisted by dynamic scattering [Figs.~\ref{Fig1}(h) and~\ref{Fig1}(i)], analogous to the dynamic scattering observed with an electric field \cite{konshina18}. At even higher input powers, the acoustic field promotes the formation of stripes within the dynamic scattering located at the acoustic pressure anti-nodes [Figs.~\ref{Fig1}(j) and~\ref{Fig1}(k)]. Lastly, applying an input power above \SI{400}{\milli\watt} induces a transition into an isotropic phase [Fig.~\ref{Fig1}(l)]. As can be seen in Fig.~\ref{FigS1}(a) in the Supplemental Material, the temperature within the channel increases linearly with the SAWs, leading to a nematic-isotropic phase transition at high input power. For low intensities, in the region of stripe patterns, however, the temperature increase is not significant. Experiments also indicate that the transmitted light intensity through acoustically induced structures increases once the system reaches the white stripe patterns, achieving maximum intensity with the colored stripe pattern, followed by a decrease in intensity in the brown stripes’ region [Fig.~\ref{FigS1}(b)].

To characterize the time scales associated with acoustically induced structures of NLCs, we first measure the time to reach a stable optical appearance after turning on the acoustic; we refer to this time as the \textit{response time}. We also measure the \textit{relaxation time}, which is the time after the acoustic field is switched off needed for the structures to relax back to the initial homeotropic dark state. Both the response and relaxation times corresponding to all structures are of the order of seconds (Fig.~\ref{FigS2}). Furthermore, the response time gradually increases when increasing the input power until it reaches a maximum for the color stripes [Fig.~\ref{FigS2}(a)]; the peak corresponds to the highest transmitted light intensity. After that, the response time decreases, reaching a plateau after brown stripes. In contrast, the relaxation time increases when increasing the input power throughout the striped patterns, reaching a plateau during the dynamic scattering regime [Fig.~\ref{FigS2}(b)]. Once the system reaches dynamic scattering with lines [Figs.~\ref{Fig1}(j) and \ref{Fig1}(k)], the relaxation time increases again until the isotropic phase is formed.

The spatial orientation of the NLC molecules in the acoustically induced structures can be resolved by performing FCPM. The observations are made along the top view (\textit{xy} plane) and cross section (\textit{yz} plane) of the channel, as shown in Figs.~\ref{Fig2}(a)-\ref{Fig2}(d). We focus on the region of stripe patterns that appear at low acoustic field intensities. High fluorescence intensity indicates that the director field is oriented parallel to the polarization of the laser beam, while a low fluorescence signal indicates an orthogonal orientation. Applying SSAWs to the NLCs induces an intense fluorescent signal in the vicinity of the acoustic pressure nodes, which indicates a tilting of the director field. Increasing the input power increases the FCPM signal, which corresponds to an even stronger tilting of the director towards the polarized laser beam. As a reference, Fig.~\ref{Fig2}(a) and sketch \ref{Fig2}(e), show the behavior of the NLC in a relaxed homeotropic state. The reorientation gradually evolves from a relatively small tilting, which corresponds to the formation of the white stripes [Fig.~\ref{Fig2}(b) and sketch \ref{Fig2}(f)], to a more pronounced tilted director, corresponding to the colored stripes [Fig.~\ref{Fig2}(c) and sketch \ref{Fig2}(g)]. At higher input power the color stripes undergo a discontinuous transition into brown stripes, whose formation results from a larger inclination of the molecules [Fig.~\ref{Fig2}(d) and sketch \ref{Fig2}(h)]. Stripe patterns collapse once the system switches into the dynamic scattering regime. The measured fluorescent signal intensities for all stripe patterns as a function of the channel width are summarized in Fig.~\ref{Fig2}(i). All the results of FCPM experiments indicate that reorientation of molecules in the pressure nodes is responsible for the observed structures.

To better understand the influence of the acoustic field on the confined nematic, we turn to continuum simulations based on the LdG theory \cite{degennes}. The total free energy of the system is expressed as a function of a tensorial order parameter $\mathbf{Q}=S(\mathbf{n}\mathbf{n}-\mathbf{I}/3)$, where $\mathbf{n}$ is a unit vector representing the nematic field, $S$ is the scalar order parameter of the nematic, and $\mathbf{I}$ is the identity tensor. Under this framework, the energy of the acoustic field is modeled as $f_A=I\cos^2(2\pi x/\lambda_x)\mathbf{k}\cdot\mathbf{Q}\cdot\mathbf{k}$, where $\mathbf{k}$ is the propagation vector, $I$ is the acoustic intensity, and $\lambda_x$ is a wavelength of applied acoustic field \cite{selinger02}. Combining the acoustic field with the hydrodynamic evolution of the confined NLC, the numerical simulations use a hybrid lattice Boltzmann method to simultaneously solve the Beris-Edwards and the momentum equations. The details of the model are summarized in the Supplemental Material. As shown in the predicted cross polarized image for low acoustic intensity in Fig.~\ref{Fig2}(j), we find that the stripe patterns arise because of the periodic distribution of the order parameter across the channel imposed by the periodicity of the acoustic wave. The acoustic wave tilts the molecules in the vicinity of the pressure nodes (see also Fig.~\ref{FigS3}) corresponding to the balance between the elastic forces of NLC and acoustic forces. The simulations indicate the onset of acoustic streaming flows [Fig.~\ref{Fig2}(k)], which have been theorized as an important phenomenon driving the alignment of molecules under SSAWs \cite{kapustina08}. Based on these results, we hypothesize that the stripe patterns become unstable and collapse into a turbulent-like flow behavior once the acoustic forces and acoustic streaming dominate over the elastic forces. While numerical predictions support experimental observation in the case of low acoustic intensity, high acoustic intensity dynamics are beyond the limitations and assumptions of our model.

\section{ACOUSTICALLY INDUCED STRUCTURES IN NEMATIC FLOW}
In this section we present results that combine the acoustic field with pressure-driven flow in the microfluidic channel. We use the Ericksen number, a relative measure of the viscous and elastic forces, as a dimensionless quantity for the flow velocity of the NLC. The Ericksen number is defined as $Er=(\gamma ul)/K$, with $\gamma$ being the rotational viscosity, $u$ being the average measured flow velocity, $l$ being the channel hydraulic diameter, and $K$ being the single elastic constant of the nematic 5CB. Similarly, we use the streaming Reynolds number as a dimensionless quantity for the acoustic wave intensity. The streaming Reynolds number relates the oscillatory forces and the viscous dissipation forces, being defined as $R_s=U_0^2/\nu\zeta$. Here, $U_0$ is the characteristic velocity of the wave, $\nu$ is the kinematic viscosity, and $\zeta$ is the frequency associated with the oscillatory flow. As shown in the Supplemental Material, $R_s \sim V^2$, where $V$ is the input voltage from the RF signal generator.

\begin{figure*}[ht!]
\includegraphics{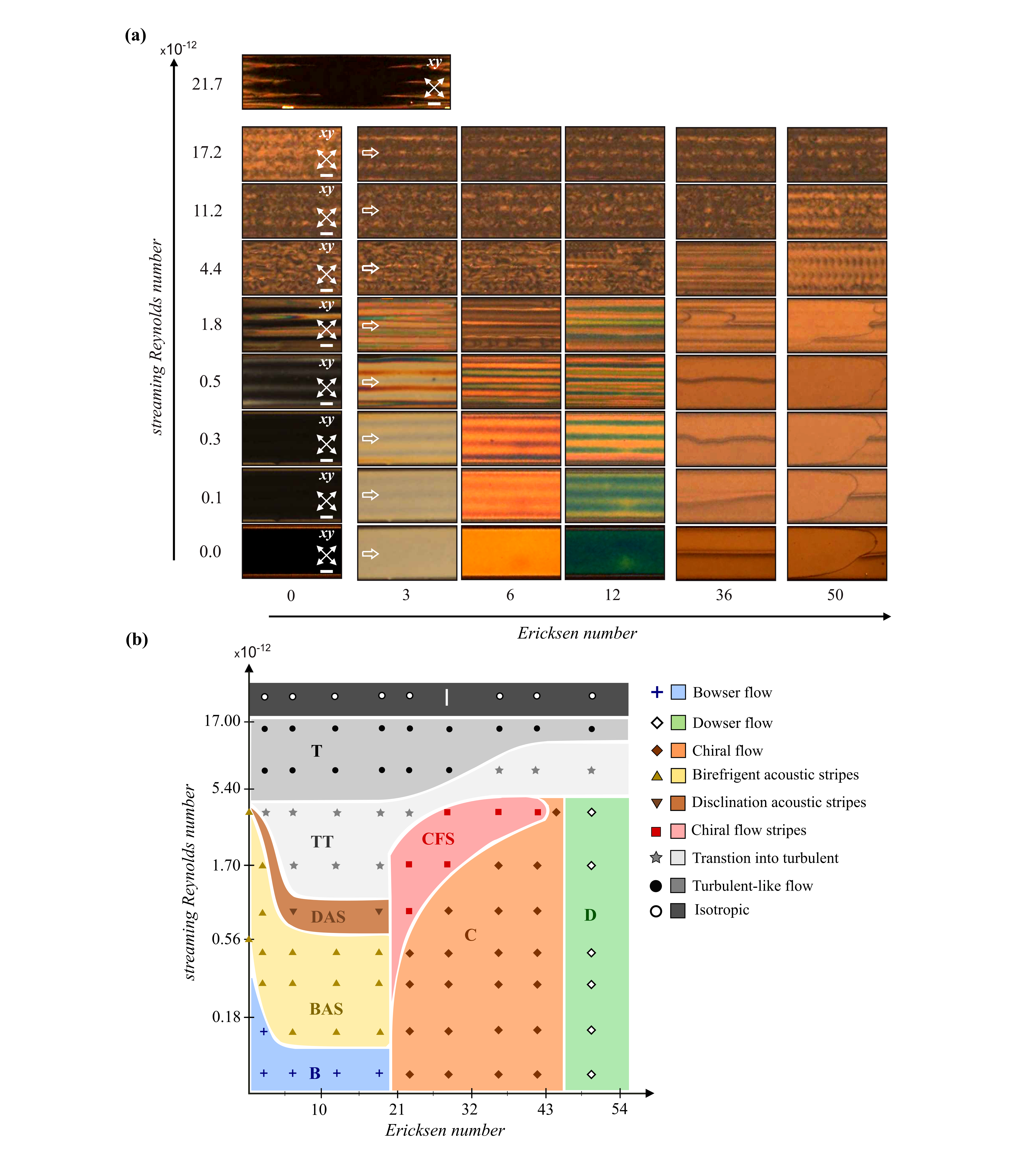}
\caption{\label{Fig3} Acoustically induced structures in nematic flow. (a) Optical responses are obtained by applying an acoustic field to the topological states of the nematic flow. Without acoustic field ($R_s = 0$), characteristic bowser, chiral and dowser states arise for $Er < 21$, $21 \leq Er < 50$, and $Er \geq 50$, respectively. In the bowser state, low acoustic intensities ($R_s < 1.8 \times 10^{-12}$) lead to stripe patterns. As the acoustic intensity is increased, the NLC transitions into a dynamic scattering regime. Chiral and dowser states dominate orientation of the director field until the acoustic intensity is not high enough ($R_s < 5.5 \times 10^{-12}$). In the range $5.5 \times 10^{-12} < R_s < 1.7 \times 10^{-11}$ the system exhibits a dynamic scattering regime due to the domination of acoustic forces. At $R_s > 1.7 \times 10^{-11}$ the NLC transitions into an isotropic phase, regardless of flow strength. White empty arrows indicate the direction of pressure-driven flow. Scale bars are \SI{100}{\micro\metre}. (b) State diagram of NLC 5CB under SSAWs and pressure-driven flow obtained from the experimental observations marked by symbols. }
\end{figure*}

We first perform experiments under pressure-driven flow for 5CB in a microfluidic channel without an acoustic field ($R_s=0$) [Fig.~\ref{Fig3}(a)]. Without flow ($Er = 0$), the channel is dark under POM due to the homeotropic alignment of the NLC. Upon starting the flow, birefringent colors start to appear, indicating a slightly bowed uniform director field towards the flow direction — known as the bowser state \cite{copar20}. The bowser state is stable only in the weak flow regime ($0 < Er < 21$), where the orientational order and flow are only weakly coupled and the structure is largely dictated by the surface anchoring. An increase in the pressure to reach the medium flow regime ($21 \leq Er < 50$) induces a continuous transformation of the bowser state into a chiral nematic state, with left and right-handed domains separated by a flexible soliton-like structure in the center of the channel \cite{copar20}. In the chiral state, the coupling of orientational order and flow leads to backflow effects. With higher flow rates ($Er \geq 50$), the nematic undergoes a discontinuous transition into a flow-aligned state known as a dowser state \cite{copar20}. A flow-aligned dowser state is observed only in the strong flow regime, in which LC molecules are primarily oriented along the main axis of the channel. All these nematic flow states have been observed and reported in previous publications and are in agreement with our observations \cite{sengupta13,copar20}.

\begin{figure*}[ht!]
\includegraphics{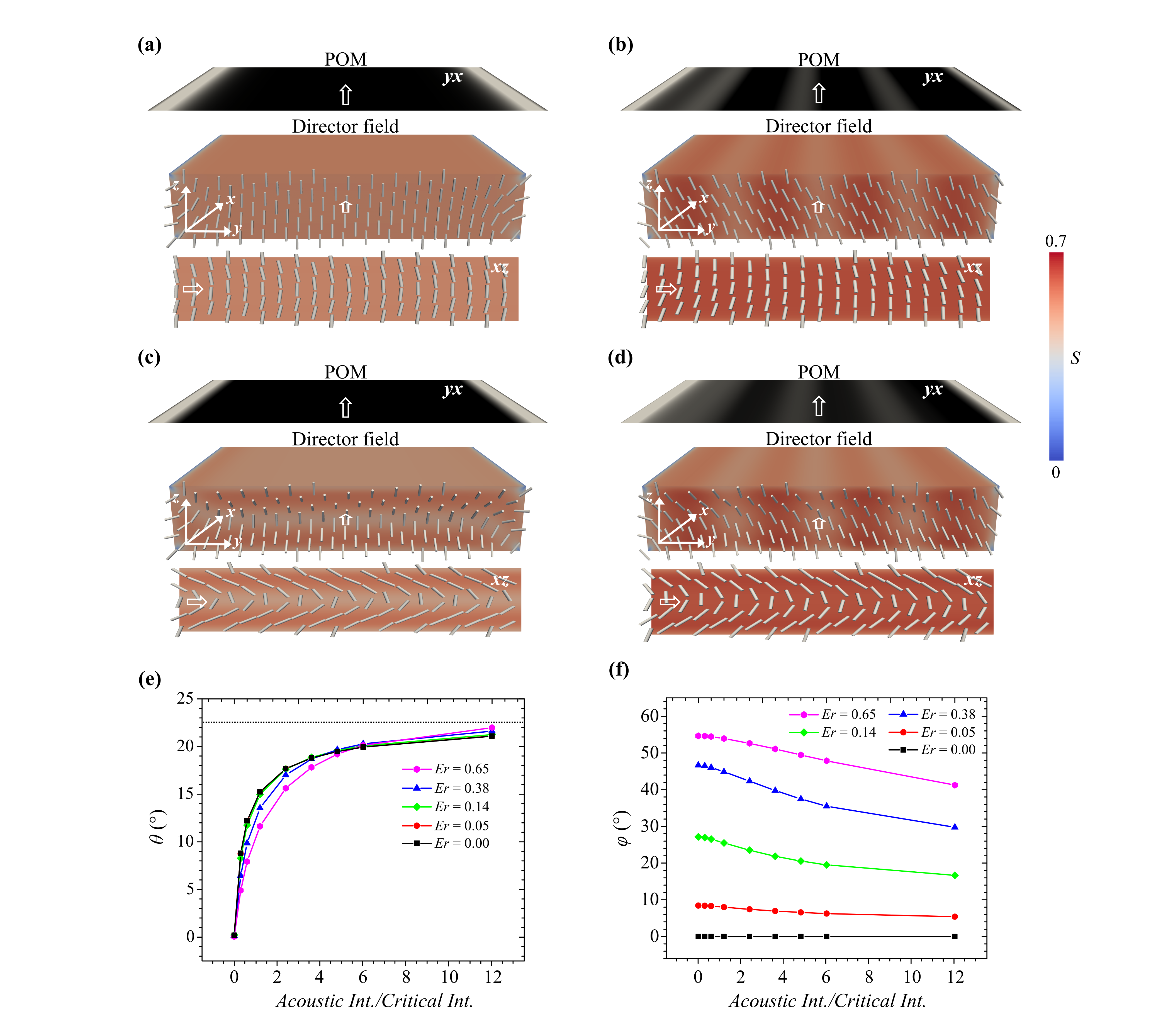}
\caption{\label{Fig4} Simulations showing the predicted molecular orientation under weak nematic flows and acoustic fields. (a)-(d) Predicted cross polarizer images, three-dimensional scalar order parameter, and orientation of director field in \textit{yz} and \textit{xz} planes of microchannel. White empty arrows indicate the direction of flow. (a) Bowser state with $Er = 0.05$ and no acoustic field. (b) Low-intensity acoustic field in a bowser state with $Er = 0.05$. (c) Bowser state with $Er = 0.65$ and no acoustic field. (d) Low acoustic field in a bowser state with $Er = 0.65$. (e) Angle $\theta$ of tilted nematic molecules across the width of the channel, measured at the acoustic pressure nodes located at $1/6$ of the maximum height. Variations in $\theta$ are observed for different flow velocities. Dotted line indicates the Rayleigh angle of incidence. (f) Angle $\varphi$ of tilted nematic molecules along the channel measured at the acoustic pressure nodes at $1/6$ of the maximum height. Variations in $\varphi$ are observed for different flow velocities. Numerical analysis with $Er$ between $0$ and $0.65$. Note that the length scales in simulations are smaller than those in experiments. }
\end{figure*}

\begin{figure*}[ht!]
\includegraphics{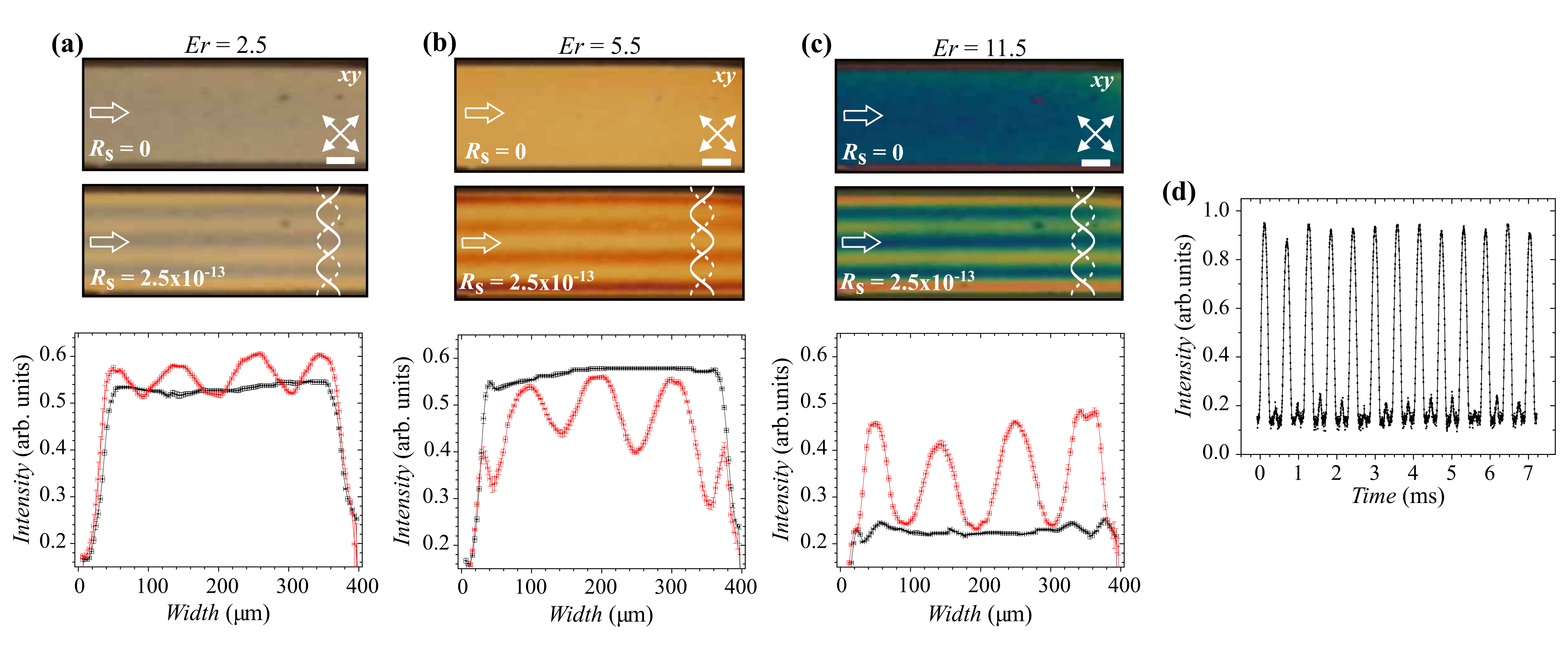}
\caption{\label{Fig5} POM images of stripe patterns in bowser state and analysis of the transmitted light intensity. (a) Applying an acoustic field with $R_\text{s} = 2.5 \times 10^{-13}$ in the bowser state with $Er = 2.5$ leads to peaks of transmitted light intensity in the vicinity of the pressure nodes of SSAWs. (b) The same acoustic field in a stronger nematic flow with $Er = 5.5$ shows peaks of transmitted light intensity at the SSAWs anti-nodes. (c) Further increasing the flow velocity to $Er = 11.5$ switches the peaks of intensity back to the acoustic pressure nodes. In graphs, black lines represent the optical response without acoustic field ($R_\text{s} = 0$), and red lines with acoustic field applied ($R_\text{s} = 2.5 \times 10^{-13}$). White empty arrows indicate the direction of the flow. Scale bars are \SI{100}{\micro\metre}. (d) Experiments show the response times for acoustically induced stripe patterns in the sub-millisecond regime, for flows in the range $0 < Er < 5.5$. }
\end{figure*}

Applying SSAWs on different topological states under flow in an orthogonal direction generates a reorientation of the director field [Fig.~\ref{Fig3}(a)]. Low SSAW input power within the range $1.8 \times 10^{-13} < R_s < 1.8 \times 10^{-12}$ leads to the formation of stripe patterns in the bowser state ($0 < Er < 21$). As the input power increases up to $R_s \sim 4.4 \times 10^{-12}$, the stripe patterns undergo a discontinuous transition to brown stripes. Higher input powers ($R_s > 5.5 \times 10^{-12}$) induce a transition into a dynamic scattering phase. In contrast with the bowser state, chiral and dowser states ($Er \geq 21$) dominate the alignment of nematic molecules when $R_s < 1.8 \times 10^{-12}$. At higher acoustic intensities, acoustically induced reorientation competes with the flow realignment until the system reaches the dynamic scattering phase. Regardless of the flow velocity, the system evolves into the isotropic phase when sufficient input power of SSAW is applied.

The acoustically induced structures observed under flow are summarized in a state diagram in Fig.~\ref{Fig3}(b). The diagram shows three distinct regions of interest. The first (B, C, and D) corresponds to the regimes where the molecular orientation of the NLC is dominated by the nematic flow, corresponding to the bowser, chiral, and dowser states, respectively. The second (T and TT) corresponds to the dynamic scattering regime, with turbulent-like flow behavior dominated by acoustic forces. The third region (BAS, DAS, and CFS) includes the stripe regimes, where the molecular orientation consists of superimposed stripe patterns, induced by acoustic forces, and the nematic flow-mediated molecular alignment of the bowser or chiral states. The isotropic state (I) is found on the top of the diagram.

Numerical simulations support the experimental observations in the bowser state in the presence of the acoustic field and a Poiseuille flow applied orthogonal to each other (Fig.~\ref{Fig4}). In the case of the bowser state with $Er = 0.05$, and in the absence of an acoustic field, the uniform homeotropic director is bowed slightly in the flow direction [Fig.~\ref{Fig4}(a)]. Applying a low acoustic intensity on the same bowser state promotes the tilting of the molecules across the width of the channel (\textit{y}-direction) in the vicinity of the acoustic pressure nodes [Fig.~\ref{Fig4}(b)]. Changes in the scalar order parameter can also be detected and visualized as stripe patterns in the simulated cross polarizer images. Without an acoustic field, an increment of the nematic flow to $Er = 0.65$ increases the bowing of the director in the direction of the flow, as shown in Fig.~\ref{Fig4}(c) by the change in the scalar order parameter. If the same acoustic field in the low intensity regime is now applied simultaneously (with $Er = 0.65$), the tilting of the molecules across the width of the channel in the vicinity of the acoustic pressure nodes is more visible than before, which corresponds to the formation of sharper stripe patterns in the cross polarizer images [Fig.~\ref{Fig4}(d)]. In Figs.~\ref{Fig4}(e) and \ref{Fig4}(f) we quantify the changes of the director orientation on the bowser state across the width (angle $\theta$) and along the length (angle $\varphi$) of the channel (Fig.~\ref{FigS4}) induced by the acoustic field. The analysis corresponds to the pressure nodes at $1/6$ of the total height of the channel, where the bowing is most pronounced according to numerical calculations spanning the entire height of the channel (Fig.~\ref{FigS5}). As seen in Fig.~\ref{Fig4}(e), increments in the acoustic intensity increase the tilting of the molecules across the width of the channel as the acoustic forces overcome the elastic forces, reaching a plateau around the Rayleigh angle $\theta_r = \sin^{-1} (v_{\text{LC}}⁄v_s) \sim 23^\circ$ \cite{selinger02}. Here, $v_{\text{LC}}\approx$ \SI{1500}{\metre/\second} and $v_\text{s}\approx$ \SI{3900}{\metre/\second} are the speed of sound in 5CB and lithium niobate, respectively \cite{grammes95,Li20}. Slight variations are visible when changing the Ericksen number in the range $0.00 < Er < 0.65$. Our simulations suggest a decrease in the tilting of the molecules along the length of the channel in the vicinity of the pressure nodes when the intensity of the acoustic field is increased (Fig.~\ref{Fig4}). As observed in Fig.~\ref{Fig4}(f), an increase in the flow velocity bows the director along the channel increasing the value of $\varphi$. In addition, increments in the acoustic intensity reduce the values of $\varphi$, and compete with the reorientation of the director field in the direction of the flow. Similar behaviors regarding angles $\theta$ and $\varphi$ are observed on the pressure nodes placed at $1/2$ of the maximum channel height (Fig.~\ref{FigS6}), although the values of $\varphi$ remain small, with no significant variations even at different velocities.

\section{OPTICAL MANIPULATION OF THE ACOUSTIC OPTOFLUIDIC DEVICE}
We next examine how the optical intensity is modified by changes to the SSAWs at a constant flow rate. POM images are shown in Fig.~\ref{Fig5}(a), where only a nematic flow with $Er = 2.5$ is applied; the optical intensity increases uniformly within the channel. For the same flow, the application of an acoustic intensity with $R_s = 2.5 \times 10^{-13}$ generates peaks of transmitted light intensity located at the pressure nodes [Fig.~\ref{Fig5}(a)]. In contrast, increasing the flow to $Er = 5.5$ while maintaining the same acoustic intensity induces a shift in the spatial location of the peaks of transmitted light intensity. Here, the peaks appear at the pressure anti-nodes, along with a reduction in the maximum value of the intensity compared with the base value obtained when only the flow is applied [Fig.~\ref{Fig5}(b)]. Furthermore, increasing the flow velocity to $Er = 11.5$ at the same acoustic intensity, shifts the location of the transmitted light intensity peaks back to the pressure nodes [Fig.~\ref{Fig5}(c)]. In this case, the intensity is significantly higher than the base value observed with only flow. Applying an alternative flow to acoustically induced stripe patterns results in sub-millisecond response times [Fig.~\ref{Fig5}(d)], indicating that the NLC responds much faster to changes induced by flow than changes induced by SSAWs (Fig.~\ref{FigS2}).

\section{DISCUSSION AND CONCLUSION}
The results of experiments ad simulations presented in this work indicate that combinations of SAWs and fluid flow enable formation of previously unknown stable nematic morphologies/states that do not exist at equilibrium. These structures, which are the combined result of hydrodynamic forces, acoustic forces, and elastic forces, exhibit sub-millisecond on/off states of brightness that can cover the entire dimensions of a microfluidic channel, or that can be localized to narrow stripes located in the corresponding pressure nodes (or anti-nodes) of the acoustic waves. Their formation can be controlled by the applied fields, and could be used to affect the spatial distribution and rate of reaction between molecules in specific regions of the material (e.g. pressure nodes) by imposing density and elasticity constrains in the NLC solvent using acoustic waves. In addition, this system can be used to spatially patterns, aggregate or actuate particles according to a range of characteristics, including density, compressibility, and size.

By increasing the complexity of the IDTs, the strategy introduced here offers the potential to control the shape, location, and frequency of the SAWs and the corresponding location of pressure nodes, paving the way for creation of optical devices with fast response times. Such devices could be further enhanced by the application of electric fields, which would offer yet another level for control of these materials.

We conclude with a word of caution regarding temperature effects in acoustically driven LCs. Specifically, care must be exercised to avoid overheating the system upon application of SAWs, thereby limiting the geometries and strength of the fields that can be used. Such temperature effects are discussed in the Supplemental Material [Fig.~\ref{FigS1}(a)].

\begin{acknowledgments}
We thank X. Li and J. A. Martinez-Gonzalez for their contributions to the pilot experiments that led to this work. The authors gratefully acknowledge the MRSEC Shared User Facilities at the University of Chicago (NSF DMR-1420709). This work made use of the Pritzker Nanofabrication Facility of the Institute for Molecular Engineering at the University of Chicago, which receives support from Soft and Hybrid Nanotechnology Experimental (SHyNE) Resource (NSF ECCS-1542205), a node of the National Science Foundation's National Nanotechnology Coordinated Infrastructure. The fluorescence confocal polarizing microscopy technique was performed in the Integrated Light Microscopy Core at the University of Chicago, which receives financial support from the Cancer Center Support Grant (P30CA014599), RRID: SCR${\_}$019197.
\end{acknowledgments}

\bibliography{acoustic}

\pagebreak
\onecolumngrid
\appendix
\section*{Supplemental Material}
\subsection{Materials and experimental procedures}
We use NLC 5CB (4-Cyano-4'-pentylbiphenyl ) (Sigma-Aldrich), which exhibits a nematic phase between 22 and 35 $^{\circ}\text{C}$. Experiments were performed within microfluidic channels having a rectangular cross section, with height $h=$ \SI{40}{\micro\metre}, width $w=$ \SI{400}{\micro\metre}, and length $L=$ \SI{12}{\micro\metre}. The channels were fabricated out of polydimethylsiloxane (PDMS) (1:10 curing agent to PDMS base; SYLGARD 184, Dow Corning) and bonded to the piezoelectric substrate after components were exposed to air plasma (Harrick Plasma, Plasma Cleaner Model PDC-001). The channel walls were chemically treated with a 0.4 wt\% aqueous solution of N-dimethyl-n-octadecyl-3-aminopropyl-trimethoxysilyl chloride (DMOAP, Sigma-Aldrich) to induce strong homeotropic surface anchoring of 5CB molecules. Interdigitated transducers (IDTs) were patterned into a 128$^{\circ}$ Y-cut of LiNbO$_3$ piezoelectric substrate (Roditi) by using standard lithography techniques and vapor deposition of \SI{10}{\nano\metre} Pt adhesion layer followed by \SI{80}{\nano\metre} Au layer. The single electrode transducer pitch was set to \SI{50}{\micro\metre} to achieve a wavelength of \SI{200}{\micro\metre}. The high voltage RF signal was generated using a GHz generator (Hewlett Packard, Model E4431B), and subsequently amplified using a power amplifier (Minicircuits, Model ZHL-1-2W-N+). Before any measurement, channels were filled with 5CB in the isotropic phase and then slowly cooled down to room temperature. The temperature was controlled using a Linkam PE120 temperature controller on a hot stage under the microscope.

\subsection{Polarized optical microscopy, fluorescence confocal polarizing microscopy, and pressure-driven flow}
Acousto-optical characterization was performed with a polarized optical microscope (POM) Olympus BX60 with 4x/10x air objectives, in transmission mode, and a fluorescence confocal polarizing microscope (FCPM) with Leica SP5 STED under 5x/10x air objectives. For FCPM imaging, 5CB was doped with 0.01 wt\% of the fluorescent dye Nile Red (Sigma Aldrich) and excited by a \SI{561}{\nano\metre} laser beam. Transmitted light intensity was measured by ImageJ. We precisely controlled and manipulated the flow by using a pressure controller-driven system (OB1, Elveflow). The flow rate was varied in the range from 0.01 to \SI{10.80}{\micro\litre/\hour}, corresponding to flow velocities ranging from 0.05 to \SI{85}{\micro\metre/\second}. The characteristic Reynolds number $Re=\rho v l/\nu$ ranged between $10^{-7}$ and $10^{-4}$, considering an effective dynamic viscosity of $\nu=$ \SI{50}{\milli\pascal s}. The material density for 5CB is $\rho=$ \SI{1.024}{\kilo
\gram/\metre^3}, and the hydraulic diameter of the rectangular microfluidic channel is estimated to be $l = 4wd/2(w + h) =$ \SI{72.7}{\micro\metre}. The corresponding Ericksen number $Er = h v l / K$, with a single elastic constant approximation $K=$ \SI{5.5}{\pico\newton}, was varied between 0.03 and 55.

\subsection{Numerical simulations}
Continuum simulations are based on the LdG formalism \cite{Lubensky95}, where the free energy is a function of the tensorial order parameter $\mathbf{Q}$
\begin{equation}\label{eq1}
\mathbf{Q}=S(\mathbf{n}\mathbf{n}-\mathbf{I}/3).
\end{equation}
In Eq.~\ref{eq1}, $S$ is the maximum eigenvalue of $\mathbf{Q}$ and $\mathbf{n}$ is the eigenvector associated with $S$. The total free energy of the NLC is defined as
\begin{equation}\label{eq2}
F=\int_V(f_{\text{LdG}}+f_{\text{el}}+f_{\text{A}})\text{d}V+\int_{\partial V}f_{\text{surf}}\text{d}S
\end{equation}
where $f_{\text{LdG}}$ is the short-range free energy, $f_{\text{el}}$ is the long-range elastic energy, $f_\text{A}$ is the acoustic energy, and $f_{\text{surf}}$ is the surface free energy due to anchoring. $f_{\text{LdG}}$ is given by \cite{degennes}
\begin{equation}\label{eq3}
f_{\text{LdG}} ={A \over 2} \left( 1 -{U \over 3} \right)  \text{Tr} \left[ \mathbf{Q}^2 \right] - {AU \over 3} \text{Tr} \left[ \mathbf{Q}^3  \right] + {AU \over 4} \left( \text{Tr} \left[ \mathbf{Q}^2 \right] \right)^2
\end{equation}
where $A$ and $U$ are phenomenological parameters and the scalar order parameter in the bulk ($S_\text{Bulk}$)
is determined by $S_\text{Bulk}= {1\over 4} + {3\over4} \sqrt{1 -{3\over 8 \ U}}$. By considering one-constant approximation, elastic energy reads \cite{Mori99}
\begin{equation}\label{eq4}
f_{\text{el}} ={L \over 2} \left( \mathbf{\nabla Q} \right)^2
\end{equation}
where $L$ is a single elastic constant. The acoustic energy is considered as follows \cite{selinger02}
\begin{equation}\label{eq5}
f_{\text{A}} = I \text{cos} ^2 \left( 2 \pi x / \lambda_x \right) \mathbf{k\cdot Q \cdot k}
\end{equation}
where $\mathbf{k}$ is the propagation vector, $I$ is the acoustic intensity, and $\lambda_x$ is a wavelength of applied acoustic field. The simulations consider homeotropic anchoring with the free energy implemented through a Rapini-Papoular expression as \cite{Pap69}
\begin{equation}\label{eq6}
f_{\text{el}} ={1 \over 2} W_H \left( \mathbf{\nabla Q} -\mathbf{\nabla Q^0} \right)^2
\end{equation}
with the surface-preferred tensorial order parameter, $\mathbf{Q}^0=S_\text{Bulk} (\nu \nu-\mathbf{I}/3)$ and normal surface $\nu$. The temporal evolution of $\mathbf{Q}$ is simulated by a hybrid lattice Boltzmann method that was used to simultaneously solve a Beris-Edwards equation and a momentum equation. The Beris-Edwards equation is given as \cite{Ber94}
\begin{equation}\label{eq7}
\left( {\partial \over \partial t} + \mathbf{u} \cdot \nabla  \right) \mathbf{Q} - \mathbf{S}=\Gamma \mathbf{H}
\end{equation}
with $\Gamma$ as the rotational diffusion constant. The tensor $\mathbf{S}$ is the generalized advection term written as
\begin{equation}\label{eq8}
\mathbf{S}=\left( \xi \mathbf{A + \Omega }\right) \cdot \left( \mathbf{Q}+{\mathbf{I} \over 3} \right) +\left( \mathbf{Q} + {\mathbf{I}\over 3} \right) \cdot \left( \xi \mathbf{A} - \Omega \right) - 2 \xi \left( \mathbf{Q} + {\mathbf{I}\over 3} \right)
\left( \mathbf{Q} : \nabla \mathbf{u} \right)
\end{equation}
where the tensors $\mathbf{A}$ and $\Omega$ are the symmetric and antisymmetric velocity gradient tensor $\nabla \mathbf{u}$, respectively. The tensor $\mathbf{H}$ is the molecular field defined as
\begin{equation}\label{eq9}
\mathbf{H}=-\left( {\delta \mathcal{F} \over \delta \mathbf{Q}} -{\mathbf{I}\over 3} \text{Tr} \left[ \delta \mathcal{F} \over \delta \mathbf{Q} \right]\right).
\end{equation}
The nematic momentum equation is \cite{Den04}
\begin{equation}\label{eq10}
\rho \left( {\partial \over \partial t} +\mathbf{u} \cdot \nabla \right) \mathbf{u}=\nabla \cdot \mathbf{\Pi} - \nu \mathbf{u}
\end{equation}
where $\mathbf{\Pi}$ is the asymmetric stress tensor known as
\begin{eqnarray}\label{eq11}
\mathbf{\Pi}&=& 2 \eta \mathbf{A} -P_0 \mathbf{I} + 2 \xi \left( \mathbf{Q} +{\mathbf{I} \over 3}  \right) \left( \mathbf{Q}:\mathbf{H} \right) - \xi \mathbf{H}\cdot \left( \mathbf{Q} +{\mathbf{I}\over 3}\right) \nonumber \\
&-& \xi \left( \mathbf{Q} +{\mathbf{I}\over 3}\right) \cdot \mathbf{H} - \nabla \mathbf{Q}: {\delta \mathcal{F} \over \delta \nabla \mathbf{Q}} +\mathbf{Q} \cdot \mathbf{H} -\mathbf{H} \cdot \mathbf{Q}
\end{eqnarray}
This hybrid Lattice Boltzmann method uses a D3Q15 grid. The parameters are chosen for an approximation of 5CB with the coherence length ($\xi_N$) as the unit length. Considering one-constant approximation, the elastic constant is $L=0.1$ with the LdG parameters $A=0.1$ and $U=3.0$. The rotational viscosity constant $\Gamma=0.133775$. The strength of homeotropic anchoring is $W_H=0.1$. The intensity of the acoustic field is in the range between $0.000\le I \le 0.020$.
The critical acoustic intensity $I_\text{crit}$ is defined as the intensity needed to reorient the director along with the acoustic wave in $zy$ plane when the flow is zero ($I=0.001$).  The simulations were done in a rectangular box with periodic boundaries condition implemented on $x$-axis and the number of simulated nodes in the grid is $N_x$=120, $N_y$=200, and $N_z$=40.

\subsection{Determination of Streaming Reynolds Number}
For a fluid of density $\rho$ and viscosity $\mu$ the Navier-Stokes equation for incompressible fluid flow may be written as
\begin{equation}
\frac{\partial \textbf{v}'}{\partial t'} - \textbf{v}' \times \omega' = - \frac{1}{\rho}\nabla \left( p' + \frac{1}{2}\rho\textbf{v'}^2 \right) + \textbf{F}' + \nu\nabla^2\textbf{v}'
\end{equation}
where $p'$ denotes pressure, $\textbf{v}'$ denotes the velocity, $\omega' = \nabla \times \textbf{v}'$ the vorticity, $\nu = \mu/\rho$ the kinematic viscosity, and $ \textbf{F}' $ a body force per unit mass. For a dimensionless analysis, we take $a$ as a characteristic length and $F_0$ a characteristic value of the force with $\zeta$ as the frequency associated with the oscillatory flow so that $U_0 = F_0/\zeta$ is a characteristic velocity of this flow. Then
\begin{equation}
\textbf{F} = \textbf{F}'/F_0, \quad  \textbf{x} = \textbf{x}'/a, \quad t = \zeta t', \quad \textbf{v} = \textbf{v}'/U_0, \quad \omega = a\omega'/U_0
\end{equation}
from which we obtain the dimensionless equation for vorticity
\begin{equation}\label{eq14}
\frac{\partial \omega}{\partial t} - \epsilon \nabla \times \left( \textbf{v} \times \omega \right)  = - \frac{\partial \Phi}{\partial t} + \frac{\epsilon}{R}\nabla^2\omega
\end{equation}
where $- \frac{\partial \Phi}{\partial t} =    \nabla \times \textbf{F}$. From Eq.~\ref{eq14} we can observe that the oscillatory flow is characterized by two dimensionless parameters: $\epsilon = U_0/\zeta a$ which is essentially an inverse of the Strouhal number, and $R = U_0 a/\nu$ a Reynolds number. No streaming induced by an oscillatory body force will occur if the body force is conservative and $R >>1$. Eq.~\ref{eq14} can be expressed as
\begin{equation}
\frac{\partial \omega}{\partial t} - \epsilon \nabla \times \left( \textbf{v} \times \omega \right)  = - \frac{\partial \Phi}{\partial t} + \frac{\epsilon^2}{R_s}\nabla^2\omega.
\end{equation}
Following the work initiated by Rayleigh, oscillatory flow induces streaming flows as a consequence of viscous attenuation close to a solid boundary, commonly associated with situations for which $k a << 1$.  Here, $k$ is the wave number and $a$ is the characteristic length. For oscillatory flows at high frequencies, it is assumed that $\epsilon << 1$. The dimensionless number $ R_s = \epsilon R$ is essentially a Reynolds number based on the velocity $\epsilon U_0$. Therefore, the magnitude of $R_s$ determines the contribution of the Laplacian of the vorticity in these flows. $R_s$ is commonly known as the streaming Reynolds number, which is used to study the development of streaming flows using acoustic waves in fluids. It relates the forces related to the oscillations and the viscous dissipation forces
\begin{equation}
R_s = \frac{U_0^2}{\nu \zeta}.
\end{equation}
In order to estimate the streaming Reynolds number from experimental data we then need values of the characteristic length $a$, the kinematic viscosity of the fluid, and a metric of the characteristic velocity of the oscillatory flow. For a square microfluidic channel, the characteristic length of the channel is given by 
\begin{equation}
a = \left( 2 w h \right)/ \left( w+h \right)
\end{equation}
where $w$ is the width of the channel and $h$ is the height of the channel that define the channel cross section. In a system using acoustic waves to generate standing acoustic waves in fluids, the value of $U_0$ relates to the amplitude of the pressure waves generated. Therefore, for a device that uses IDTs to create SAWs, $U_0$ depends on the voltage applied to the transducers and therefore to the applied power from the radio frequency source. The task now is to connect the applied power and the voltage signal to the velocity of the propagating wave in the fluid. Based on the quasi-static method for non-reflective single-electrode transducers, the total wave amplitude that exits the transducer port is defined as
\begin{equation}
\phi_s(\zeta) = \sum_{m=1}^{m} \phi_{sm}(0,\zeta) = V E(\zeta) \sum_{m=1}^{m} \hat{P}_{m}exp(-jkx_m)
\end{equation}
where it is assumed that all variables have harmonic dependence so that they are proportional to $exp(j \zeta t)$. Here $V$ is the applied voltage, $M$ is the total number of electrodes in the transducer, $k= \zeta/\gamma$ is the wavenumber with phase velocity $\gamma$, and $\hat{P}_{m} = 0, 1, 0, 1, 0, 1, ...$ is the electrode polarity for a single-electrode transducer. $E(\zeta)$ here is an element factor, which varies slowly with $\zeta$ and is often consider to be constant. For a single electrode transducer, the center frequency $\zeta_{s0}$ occurs when the electrode pitch $p_e$ equals $\lambda/2$, giving $\zeta_{s0} = \pi \gamma/ p_e $. At this frequency it is found that $E( \zeta_{s0}) = 1.694 j (\Delta \gamma / \gamma)$, getting $ E(\zeta) \approx  E(\zeta_{s0})$ for frequencies near to $\zeta_{s0}$. Defining an array factor as
\begin{equation}
A(\zeta) =  \sum_{m=1}^{m} \hat{P}_{m}exp(-jkx_m)
\end{equation}
gives $ \phi_s/V =  E(\zeta) A(\zeta)$. Of note, the transducer response $H_t(\zeta)$ is defined by the expression \cite{morgan}
\begin{equation}
H_t(\zeta) =   -j E(\zeta) A(\zeta) \sqrt{ \zeta W \epsilon_{\infty}/(\Delta \gamma / \gamma)}
\end{equation}
so that the potential of the wave generated can be written as
\begin{equation}\label{eq21}
\Phi_s/V =   - j H_t(\zeta) \sqrt{(\Delta \gamma / \gamma)/(\zeta W \epsilon_{\infty})}.
\end{equation}
Here, $W$ is the width of the electrode fingers and $ \epsilon_{\infty}$ is defined as the capacitance of a unit-aperture single electrode transducer per period. When a voltage $V$ is applied, the power absorbed by the transducer is \cite{morgan}
\begin{equation}
P_a = G_a(\zeta)\lvert V \rvert^2 /2
\end{equation}
where the conductance $G_a(\zeta)$ and the susceptance $B_a(\zeta)$ are part of the parallel elements added to the capacitance $C_t$ to determine the electrical admittance $Y_t(\zeta)$
\begin{equation}
Y_t(\zeta) = G_a(\zeta) + j B_a(\zeta) + j\zeta C_t.
\end{equation}
The power of a wave with surface potential $\phi_s$ can be shown to be \cite{morgan}
\begin{equation}
P_s = \frac{1}{4}\zeta W  \epsilon_{\infty}\lvert \phi_s \rvert^2 /(\Delta \gamma / \gamma)
\end{equation}
which leads together with Eq.~\ref{eq21} to the simple relation
\begin{equation}
G_a(\zeta) = \lvert H_t(\zeta)  \rvert^2.
\end{equation}
The fundamental response of a uniform transducer has
\begin{equation}
G_a(\zeta)  \approx G_a(\zeta_0) \left[ sin\left(X \right)/X \right]^2
\end{equation}
where $ X \equiv N_p \theta = N_p (\zeta - \zeta_0)/\zeta_0$ and $N_p = M/2$ for single-electrode transducers \cite{morgan}. Now, it can be shown that for single-electrode transducers
\begin{equation}\label{eq27}
G_a(\zeta_0)  = \alpha \zeta_0 \epsilon_{\infty}WN_p^2 (\Delta \gamma / \gamma),  \quad \alpha = 2.87.
\end{equation}
Therefore, it follows from this development that as $\zeta \rightarrow \zeta_0$
\begin{equation}
P_{s0}  =  \frac{1}{4} \alpha \zeta_0 \epsilon_{\infty} W N_p^2 (\Delta \gamma / \gamma) V^2.
\end{equation}
Eq.~\ref{eq27} relates the total voltage applied from the radio-frequency source to the power of the acoustic wave leaving an IDT that is operating at its resonance frequency $\zeta_0$. Assuming, for simplicity, that there are no losses in the system due to diffraction and dampening, we can relate the power of the traveling surface acoustic wave with the acoustic intensity of the leaking surface acoustic wave in the fluid. For the traveling wave within the fluid, we have that the acoustic intensity is related with the amplitude of the velocity profile by
\begin{equation}
I = \rho_o c u^2
\end{equation}
where $\rho_o$ is the fluid density, $c$ is the speed of sound, and $u$ is the fluid velocity related with the oscillatory motion of the fluid. Hence, we can relate the power of the acoustic wave traveling within the fluid with the characteristic velocity through a cross section area $A_o$ by
\begin{equation}
P_f = A_o \rho_o c U_0^2.
\end{equation}
Assuming that the power of the acoustic wave exiting the transducer is the power of the traveling wave within the fluid, $P_s = P_f$, we can then estimate the characteristic velocity of the wave generated by the single-electrode IDT to be
\begin{equation}
U_0^2 \approx  \frac{1}{4 A_o \rho_o c} \alpha \zeta_0 \epsilon_{\infty} W N_p^2 (\Delta \gamma / \gamma) V^2.
\end{equation}
Following this treatment, we conclude that the streaming Reynolds number $R_s$ can be estimated as a function of the voltage applied to the transducers
\begin{equation}
R_s =  \chi  V^2
\end{equation}
where 
\begin{equation}
\chi =  \frac{1}{8 \nu} \frac{ \alpha \epsilon_{\infty} W N_p^2 (\Delta \gamma / \gamma)}{A_o \rho_o c}
\end{equation}
is a constant defined by the properties of the fluid, the dimensions of the microfluidic channel, the piezoelectric properties of the material where the IDTs are located, and the architecture of the transducers.
\pagebreak
\subsection{Supplemental Figures}

\renewcommand{\thefigure}{S\arabic{figure}}

\setcounter{figure}{0}

\begin{figure*}[h]
\includegraphics{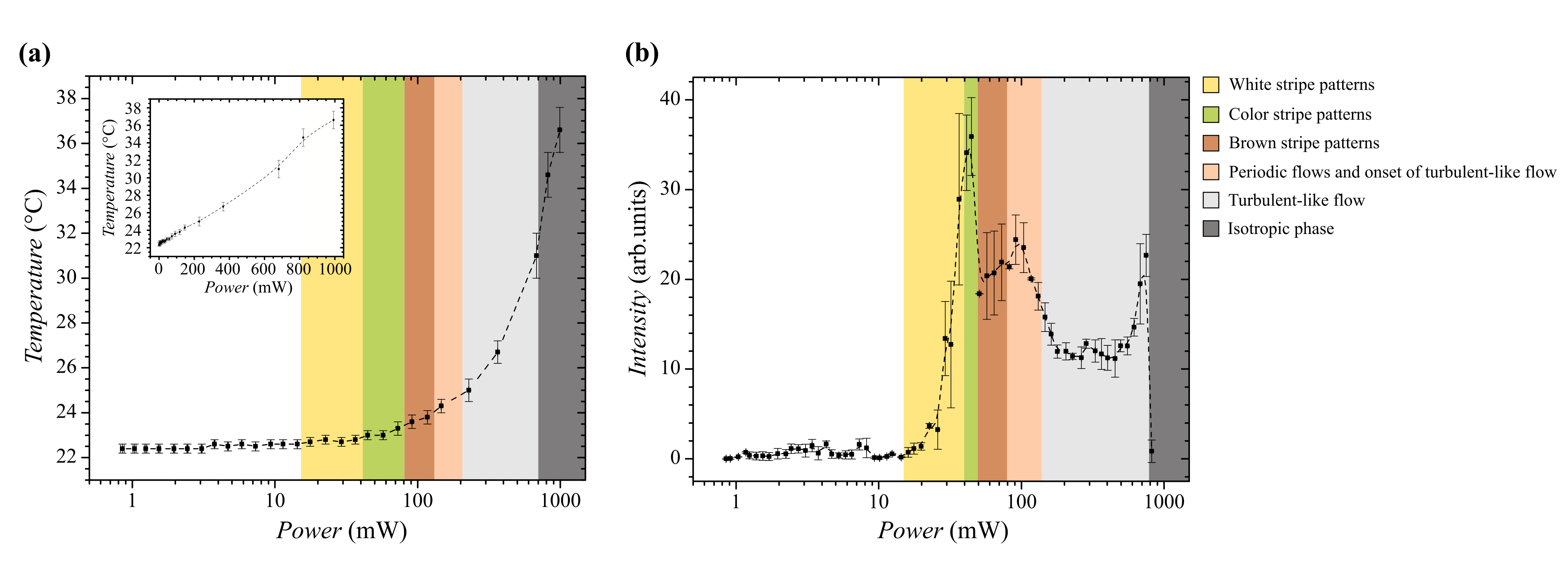}
\caption{\label{FigS1} Temperature and transmitted light intensity of NLC under SSAWs. (a) Change in temperature of a nematic in a microfluidic channel as a function of input power of RF signal (inset: linear plot). In the region of white and color stripe patterns, the temperature is not significantly elevated whereas the turbulent-like flow regime significantly increases the temperature of the system. (b) The transmitted light intensity of acoustically induced structures with the peak in the region of color stripe patterns. Color legend represents a specific structure region. }
\end{figure*}

\begin{figure*}[h]
\includegraphics{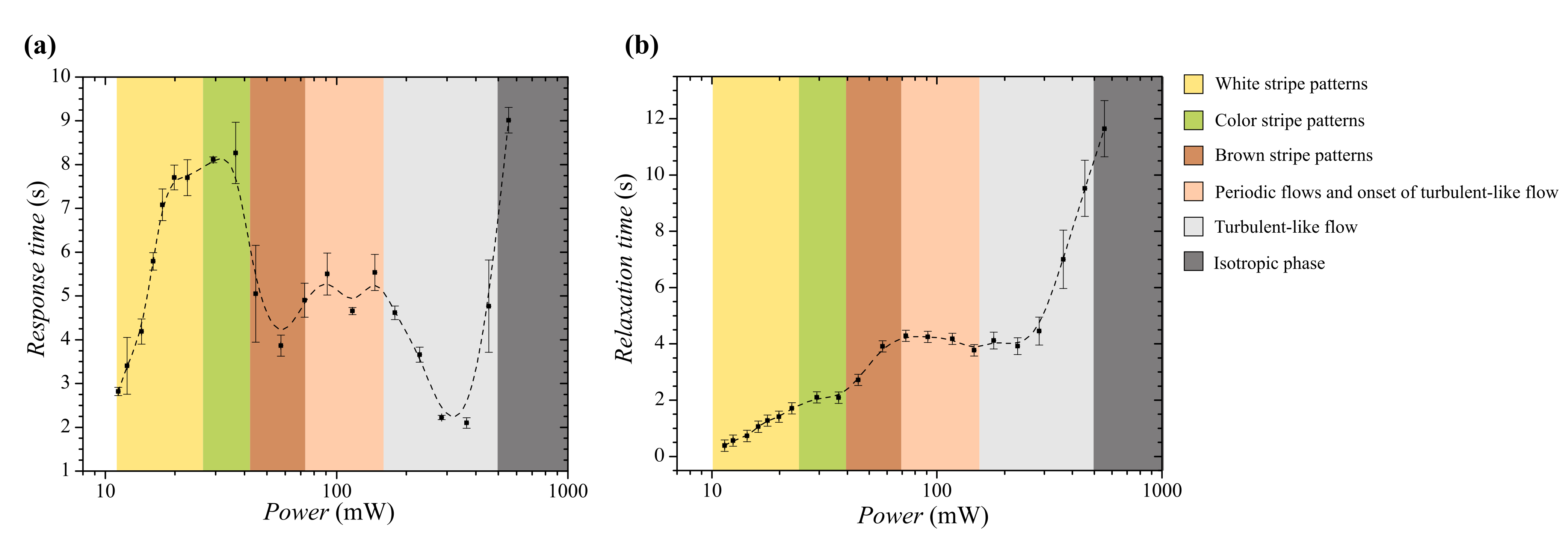}
\caption{\label{FigS2} Time scales of acoustically induced structures of NLCs. (a) Response and (b) relaxation time for the nematic to reach a steady state after turning the SSAWs on and off, respectively. Color legend represents a specific structure region. }
\end{figure*}

\begin{figure*}[h]
\includegraphics{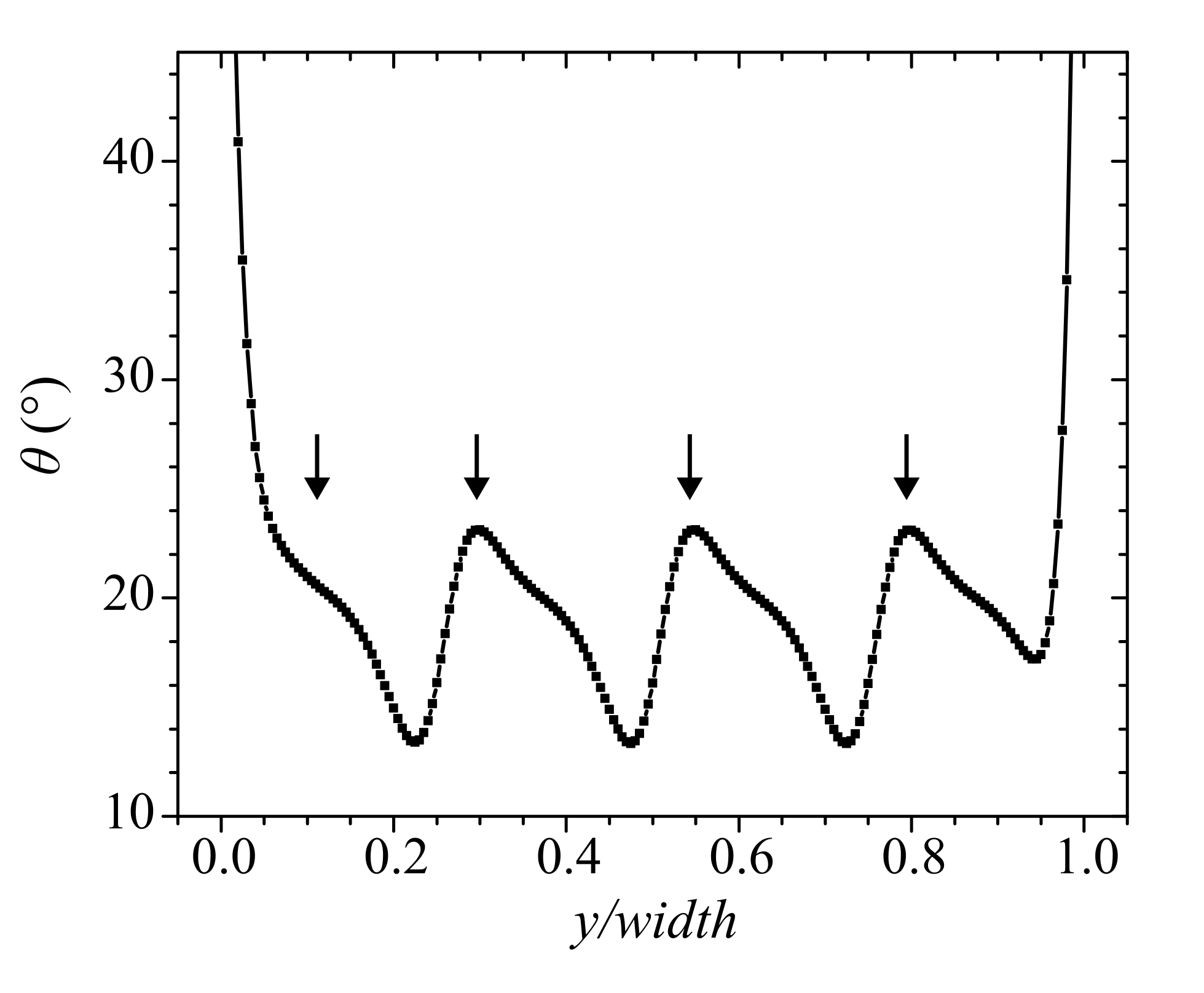}
\caption{\label{FigS3} Numerically predicted angle $\theta$ of the director field across the channel at $1/6$ of the maximum channel height under the acoustic field in Fig.~\ref{Fig2}(j). Black arrows indicate regions of pressure nodes. }
\end{figure*}

\begin{figure*}[h]
\includegraphics{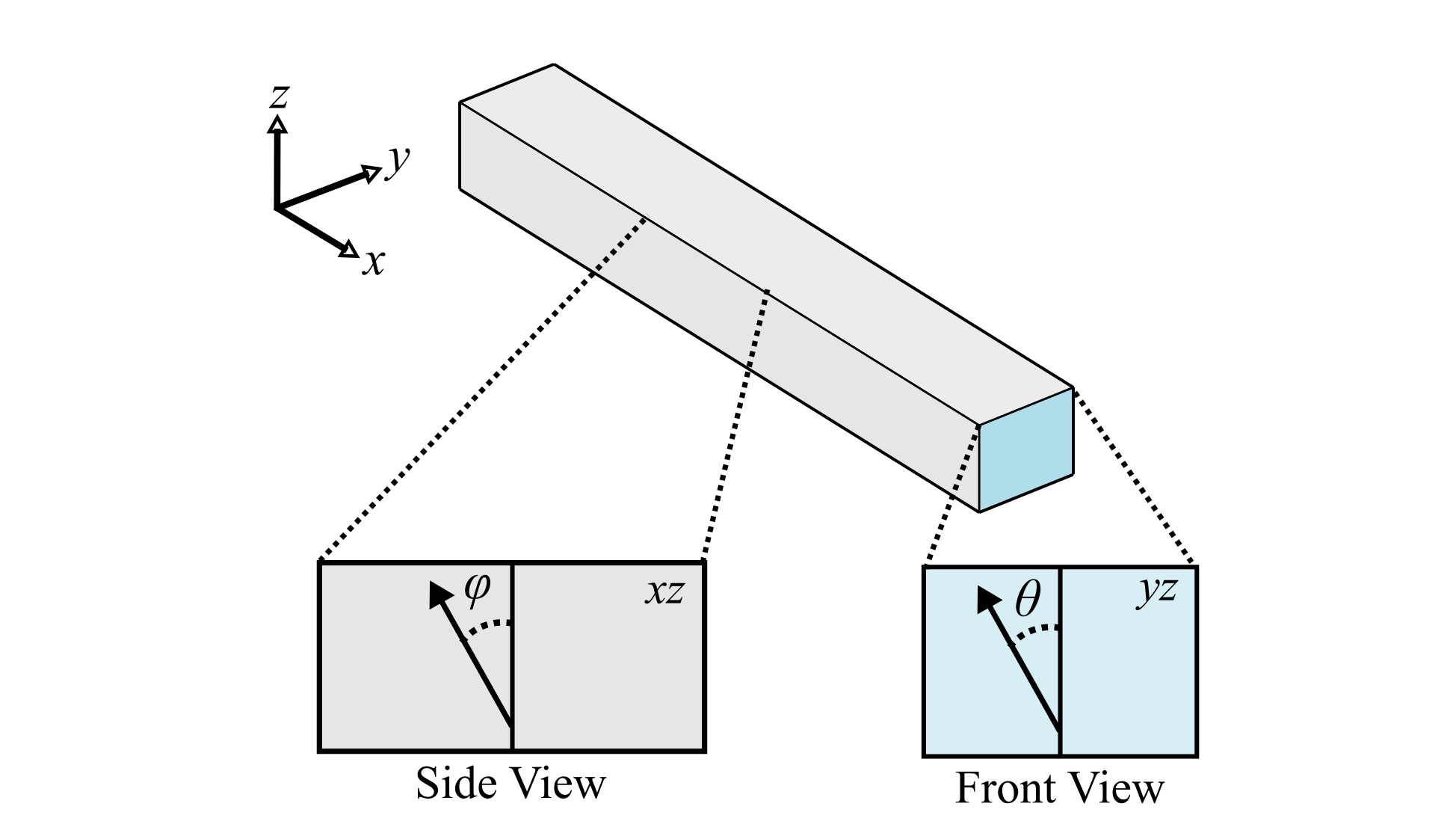}
\caption{\label{FigS4} Schematic sketch of the director orientation changes across (angle $\theta$) and along (angle $\varphi$) the microfluidic channel. }
\end{figure*}

\begin{figure*}[h]
\includegraphics{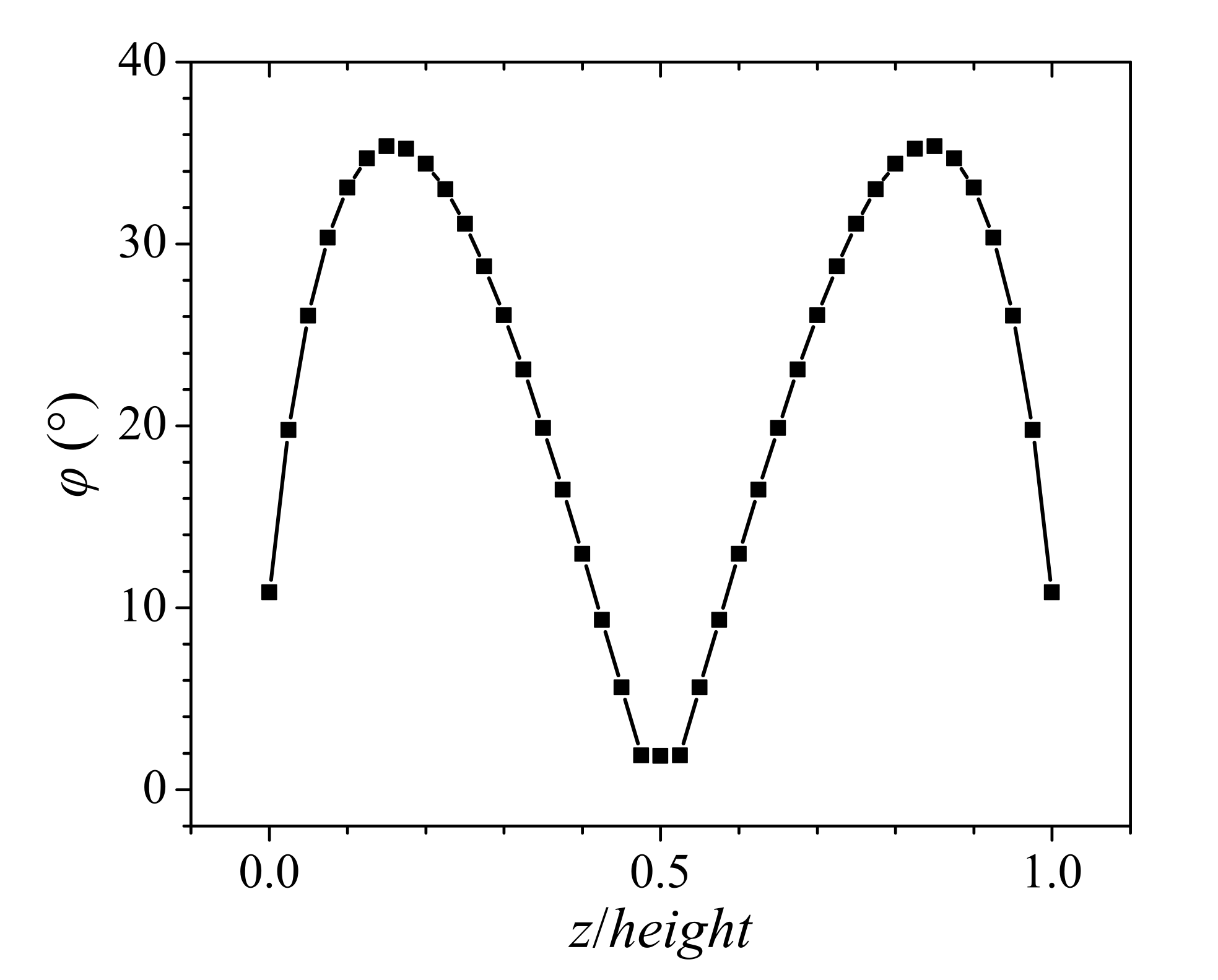}
\caption{\label{FigS5} Angle $\varphi$ of the director field in the dowser state of the weak flow regime through the height of the channel. In the weak flow regime, the nematic molecules are bowed in the direction of flow, closer to the walls more and in the center of the channel less. }
\end{figure*}

\begin{figure*}[h]
\includegraphics{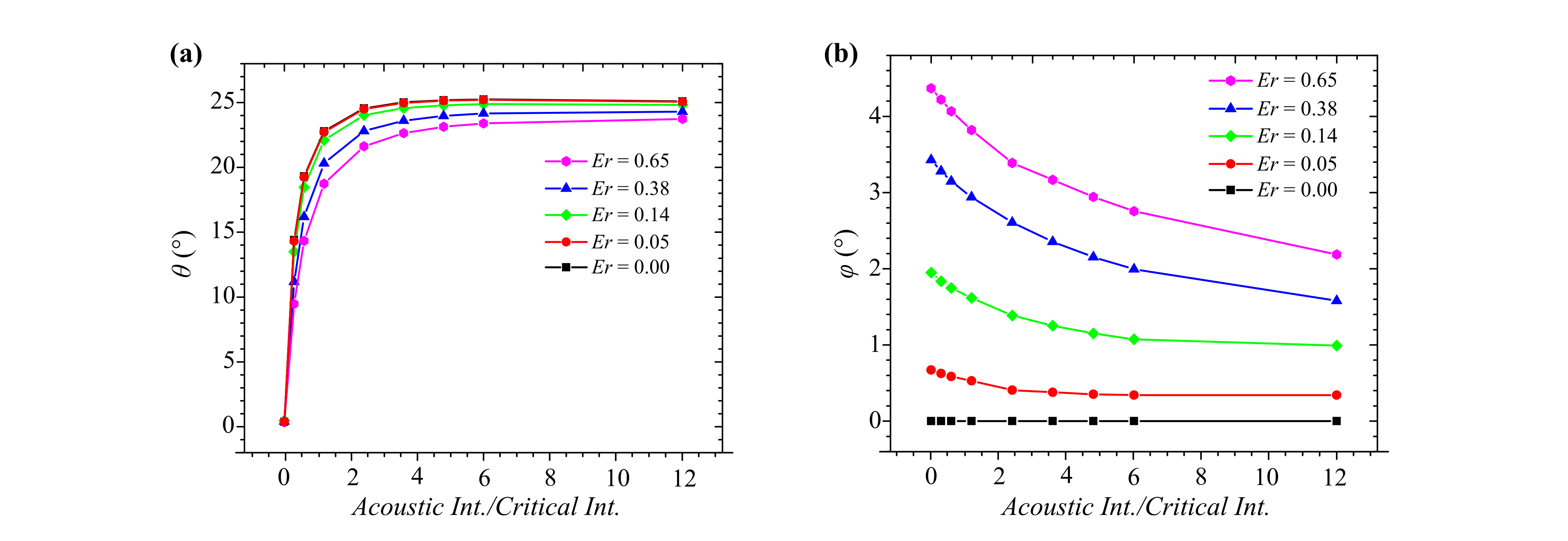}
\caption{\label{FigS6} Quantification of the director field angle changes on acoustic pressure nodes across (angle $\theta$) and along (angle $\varphi$) the channel at $1/2$ of the maximum channel height induced by the acoustic field in the bowser state. (a) Molecules tilt across the channel more at a higher acoustic intensity and less at the same acoustic intensity with a higher nematic flow. (b) Stronger flow bows the molecules more along the channel whereas higher acoustic intensity decreases the bowed shape. Numerical analysis is done in the weak flow regime with $Er$ between 0 and 0.65. Note that the scale in simulations is orders of magnitude smaller than that in experiments. }
\end{figure*}

\end{document}